\documentclass [twocolumn]{aastex62}

\newcommand \kep {\textit{Kepler}}
\newcommand \gaia {\textit{Gaia}}
\newcommand \vtot {V_\mathrm{tot}}
\newcommand \kms {\mathrm{km~s}^{-1}}
\newcommand \feh {[\mathrm{Fe}/\mathrm{H}]}
\newcommand \mh {[\mathrm{m}/\mathrm{H}]}
\newcommand \teff {T_\mathrm{eff}}
\newcommand \rp {R_\mathrm{p}}

\shorttitle{Small Planets in the Galactic Context}
\shortauthors{Bashi \& Zucker}

\begin{document}

\title{Small Planets in the Galactic Context: Host Star Kinematics, Iron, and $\alpha$~Element Enhancement}

\correspondingauthor{Dolev Bashi} \email{dolevbas@mail.tau.ac.il}

\author[0000-0002-9035-2645]{Dolev Bashi}
\affiliation{Department of Geophysics, Raymond and Beverly Sackler Faculty of Exact Sciences,\\
Tel Aviv University, Tel Aviv, 6997801, Israel}

\author[0000-0003-3173-3138]{Shay Zucker}
\affiliation{Department of Geophysics, Raymond and Beverly Sackler Faculty of Exact Sciences,\\ 
Tel Aviv University, Tel Aviv, 6997801, Israel}

\begin{abstract}
We explored the occurrence rate of small close-in planets among \kep\ target stars as a function of the iron abundance and the stellar total velocity $\vtot$. We estimated the occurrence rate of those planets by combining information from LAMOST and the California-\kep\ Survey (CKS) and found that iron-poor stars exhibit an increase in the occurrence with $\vtot$ from $f < 0.2$ planets per star at $ \vtot < 30\ \kms$ to $f \sim 1.2$ at $\vtot > 90\ \kms$. We suggest this planetary profusion may be a result of a higher abundance of $\alpha$~elements associated with iron-poor, high-velocity stars. Furthermore, we have identified an increase in small planet occurrence with iron abundance, particularly for the slower stars ($ \vtot < 30\ \kms$), where the occurrence increased to $f \sim 1.1$ planets per star in the iron-rich domain. Our results suggest there are two regions in the $(\feh,[\alpha/\mathrm{Fe}])$ plane in which stars tend to form and maintain small planets. We argue that analysis of the effect of overall metal content on planet occurrence is incomplete without including information on both iron and $\alpha$-element enhancement.  
\end{abstract}


\section{Introduction} \label{sec:intro}
Planet occurrence rate plays a crucial role in constraining theories of planet formation and evolution. One of the first analyses of planet occurrence rate has revealed, for example, that metal-rich stars are much more likely to harbor close-in giant planets \citep{Santos04,FischerValenti05}. This observation was one of the main pieces of evidence supporting the core-accretion planet formation model, as protoplanetary disks having a higher content of metals and thus enhanced surface densities of solids, promote the formation of heavy element cores of giant planets \citep{Pollack1996}. Viewed in this context, we might also expect that metal-rich disks should form terrestrial planets more efficiently than metal-poor disks, or in other words, metal-rich stars should host more planets. The launch of NASA's \kep\ space telescope has allowed us to better address these issues. \kep\ had monitored the brightness of thousands of stars over a wide field, seeking evidence for planetary transits. Great progress was achieved in the analysis of planet occurrence with \kep\ data, including progress toward estimating the fraction of Sun-like stars that harbor Earth-like planets \citep{Youdin2011, Foreman-Mackey14, Burke15} or the discovery of a radius gap between super-Earth and sub-Neptune planets \citep{Fulton17}.

In the preparation of the \kep\ Input Catalog (KIC) by \cite{Brown11} a prior based on the metallicities of nearby stars was placed on the \kep\ field star metallicities. This eventually resulted in lack of reliable metallicity estimates for a representative sample of \kep\ stars, fundamentally limiting the quality of initial \kep\ metallicity studies. While \kep\ did provide a large sample of planets for planet-metallicity studies, a follow-up spectroscopic survey was still needed in order to measure the host star metallicities. By measuring the metallicities of $152$ stars harbouring $226$ planets, \cite{Buchhave12} observed that while planets larger than $4R_\earth$ tend to orbit metal-rich stars, smaller planets are found in a wide range of metallicities.

The Large Sky Area Multi-Object Fiber Spectroscopic Telescope, LAMOST \citep{cui12}, is an ideal instrument for follow-up spectroscopic observations on \textit{Kepler} stars \citep{ZongLK2018}. The LAMOST survey provides low-resolution optical spectra for a very large sample of stars \citep{Wang96}, and recent studies have already used information gained by LAMOST spectra on the stellar iron content to study its effect on planetary occurrence. \cite{petigura2018california}, for example, have calculated the occurrence rates for different populations of close-in ($P < 100$~days) planets in several iron abundance intervals, suggesting that in most cases the occurrence rate increases with stellar metal content.

Although iron content ($\feh$) is commonly used as a proxy for overall metallicity ($\mh$), iron is undoubtedly not the only abundant refractory element in stellar atmospheres. There are other rather abundant elements with condensation temperatures comparable to iron \citep{Lodders2003, Adibekyan12}, which are very important contributors to the composition of dust in planet forming regions \citep[e.g.,][]{Gonzalez2009, Helled14}. In general, using iron abundance as a proxy to overall metallicity assumes that the abundances of all metals change proportionally to iron content \citep{Bertelli94}. However, this assumption is not always valid, especially for iron-poor stars that are enhanced in $\alpha$~elements, i.e., Mg, Si, Ti, Ca etc. \citep[e.g.,][]{Bensby14, Kordopatis15}. Consequently, there is some inconsistency in different metallicity surveys where some use $\feh$ as overall metallicity while others use $\mh$ \citep[see also,][]{Adibekyan19}.

Focusing on iron-poor stars, \cite{Adibekyan12} have noted that the frequency of planet-host stars is higher among titanium-enhanced stars. In a recent paper, \cite{BrewerMulti18} have shown that compact planetary systems tend to occur more frequently around relatively metal-poor stars. Both studies have argued that stars of lower metallicity and higher $[\mathrm{Ti}/\mathrm{Fe}]$ or $[\mathrm{Si}/\mathrm{Fe}]$ and in general higher abundance of $\alpha$~elements, are older and probably members of the chemically defined Galactic thick-disk population. However, the dependence of planet occurrence rate on stellar elemental abundances or galactic context is far from being completely understood.

The properties of the local thick disk have been characterized by many previous spectroscopic studies. 
It is commonly assumed to be older, kinematically hotter, and more metal-poor and $\alpha$-rich than the thin disk \citep[e.g.,][]{Gilmore89, Reddy03, Bensby14, Kordopatis15, BuderGALAH18, DuongGALAH18}. In the literature, there is currently no unique way to affiliate stars to either the thin or thick discs. Often, kinematic criteria are used, but classifications based on chemical composition, stellar age, spatial position or a combination thereof can also be found. It is likely that distinguishing the thin-disk population purely on the basis of certain specific criteria would result in contamination of thick disk stars. Thus, some stars with thick disk chemistry have thin disk kinematics and vice versa \citep{Bensby14, DuongGALAH18}.

Recently, in an attempt to model planet structure and composition from stellar element abundance, \cite{Santos17} have shown that disks around stars from different Galactic populations seem to be forming rocky planets with significantly different iron-to-silicate mass fractions. Their research has demonstrated that planets can form in both iron-rich thin disk stars as well as in iron-poor and $\alpha$-enhanced thick disk stars. Consequently, they have argued that their results may have impact on our understanding of the occurrence rate of planets in the Galaxy.

In this work, we aim to examine planet occurrence rate in the context of iron~content and kinematics.
In Section \ref{sec:style} we describe the way we complied our planet and stellar samples. In Section \ref{sec:methodology} we present our methodology which includes calibration of the LAMOST $\feh$ and RVs (\ref{subsec:Calibration}), estimating the stellar kinematics and planet occurrence rate (\ref{subsec:kinematics}, \ref{subsec:Occurrence}) and a practical demonstration of the close relation between stellar kinematics and $\alpha$ enhancement (\ref{subsec:kinematicsandalpha}). We show our planet occurrence rate results in section \ref{sec:Results}, and discuss our findings in section \ref{sec:Discussion}.   

\section{The Sample} \label{sec:style}
In order to study planet occurrence we need both a sample of planet-host stars and a homogeneous and unbiased sample representing the parent stellar population. Our planet-host sample is based on the California-\kep\ Survey (CKS) catalog of $1305$ \kep\ Objects of Interest (KOIs). The CKS sample has a high purity (i.e., low false-positive rate) due to intensive and rigorous vetting of false positives \citep{petigura2017california}. Recently, using \gaia\ parallaxes, \kep\ photometry, and temperatures from CKS, \cite{FultonPeti18} further constrained the range of stellar properties in their sample and finally obtained a well-defined sample of $907$ planet candidates.

We have constructed the sample of parent stellar population using \kep\ stars with known $\feh$ from the KIC \citep{Mathur2017} that are also listed in the LAMOST DR4 catalog. We cross-matched LAMOST DR4 with the KIC by identifying LAMOST spectra taken within $1\arcsec$ of a KIC star. We further restricted our sample to stars which had positive \gaia\ parallaxes, with a relative uncertainty smaller than $20\%$ \citep{GaiaDR2}. This resulted in a sample of $58206$ stars.

As stars evolve, the measured surface abundances of heavy elements show a pronounced decrease \citep{Souto18}, which might introduce unwanted bias on any analysis of the influence of those elements. Therefore, following \citet{petigura2018california} and \citet{FultonPeti18}, we limited our analysis to unevolved (FGK) stars by restricting the sample to stars with $4700\ \mathrm{K} < \teff < 6500\ \mathrm{K}$ and $\log g > 4$ that are brighter than $G = 14.4$ (assuming $K_\mathrm{P} \sim G - 0.2$\ mag for G2 stars.). We further adopted filters proposed by \citet{Lindegren18} to reduce contamination from binary stars, calibration problems or spurious astrometric solutions. After applying these cuts, we were left with a sample of $21227$ stars representing the parent stellar population.

For our planet-host sample we cross-matched our LAMOST-KIC stellar sample with that of KOI-host stars listed in \citet{FultonPeti18}, resulting in a $561$ planets orbiting $376$ different planet-host systems.

\begin{figure}[ht!]
\epsscale{1.2}
\plotone{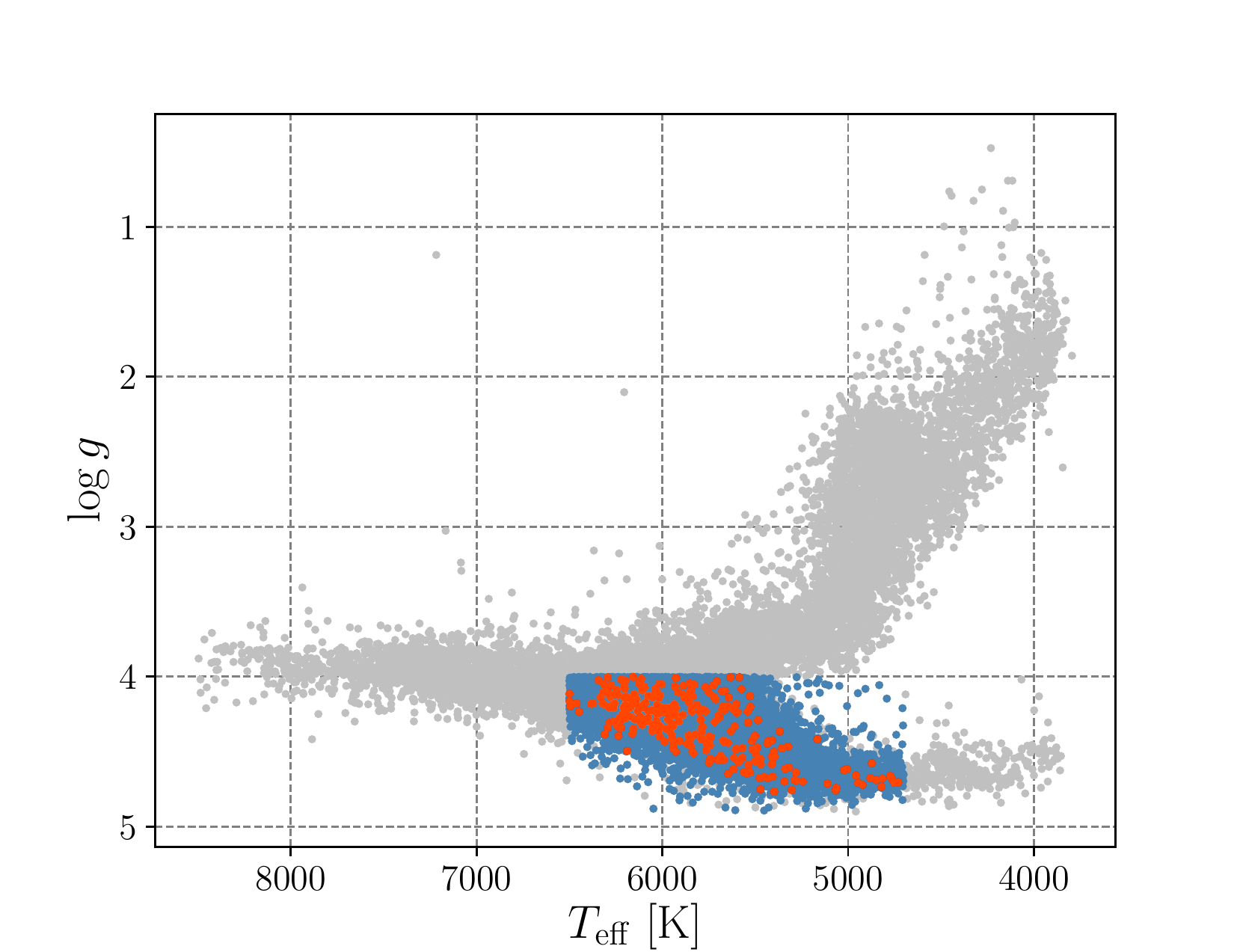}
\caption{$\log g$ as a function of effective temperature $\teff$ for the overall sample (gray), KICs that comply with our filters (blue) and the planet-host KICs listed in the CKS catalog (red).}
\label{fig:Main}
\end{figure}

We constrained the KOI sample to include only small close-in planets, i.e. $\rp = 1$\,--\,$4 R_\earth$ and $P<100$~days. The upper limit of $\rp$ was based on that of sub-Neptune size planets, while the lower limit was meant to exclude the region of low completeness, following \citet{FultonPeti18}. The final planet sample contained $391$ KOIs. Figure~\ref{fig:Main} presents (in gray) our LAMOST-KIC sample in the $(\log g, \teff)$ plane. The stars that passed our filters are marked in blue, and the planet-host stars among them in red.\

\section{Methodology} \label{sec:methodology}

\subsection{Calibration of LAMOST $\feh$ and RVs}
\label{subsec:Calibration}
The CKS and LAMOST surveys used different spectra, line lists, and model atmospheres to estimate iron abundances and heliocentric line-of-sight velocities. We therefore applied corrections, following \citet{petigura2018california}, to account for the systematic offsets. We identified $492$ stars that were listed in both CKS and LAMOST DR4, and used them to fit for linear calibrations between LAMOST $\feh$ and RV to those of CKS. Similarly to \citet{petigura2018california}, before fitting we removed seven and eight outliers where the CKS and LAMOST $\feh$ and RV differed by more than $0.2$ and $20\ \kms$, respectively. These are probably rare cases of failure in the LAMOST pipeline.

Comparing our iron abundance corrections to those used by \citet{petigura2018california}, we do find an agreement, however, it is important to note that while we have used LAMOST DR4, \citeauthor{petigura2018california} have used LAMOST DR3, and therefore it was important to repeat the calibration process.

\subsection{Stellar Kinematics}
\label{subsec:kinematics}
Stellar kinematics is the observational description of the positions and motions of the stars in the Galaxy. Combining sky coordinates, proper motions and distance estimates (from parallaxes), acquired by \gaia\ DR2 \citep{GaiaDR2}, together with RV measurements from LAMOST DR4, can provide the information required to obtain full space motions. 

It is common to present the velocity vector thus obtained in the Local Standard of Rest (LSR) frame, which is defined as the reference frame of a star at the location of the Sun, assuming this star moves in the gravitational Galactic potential in a circular orbit \citep[e.g.,][]{Coskunoglu11}. The components of this vector are usually denoted by $U$, $V$ and $W$, where $U$ is the velocity component directed outwards from the Galactic center, $V$ is aligned with the local direction of the Galactic rotation, and $W$ is positive towards the North Galactic Pole.

As for corrections for the motion of the Sun with respect to the LSR, we adopted the values $(U_\sun,V_\sun,W_\sun)=(8.50, 13.38, 6.49)\ \kms$ \citep{Coskunoglu11}. We have computed the velocity vector $(U,V,W)$ following the procedure described in \cite{JohnsonSoderblom1987}. 

\begin{figure}[ht!]
\epsscale{1}
\plotone{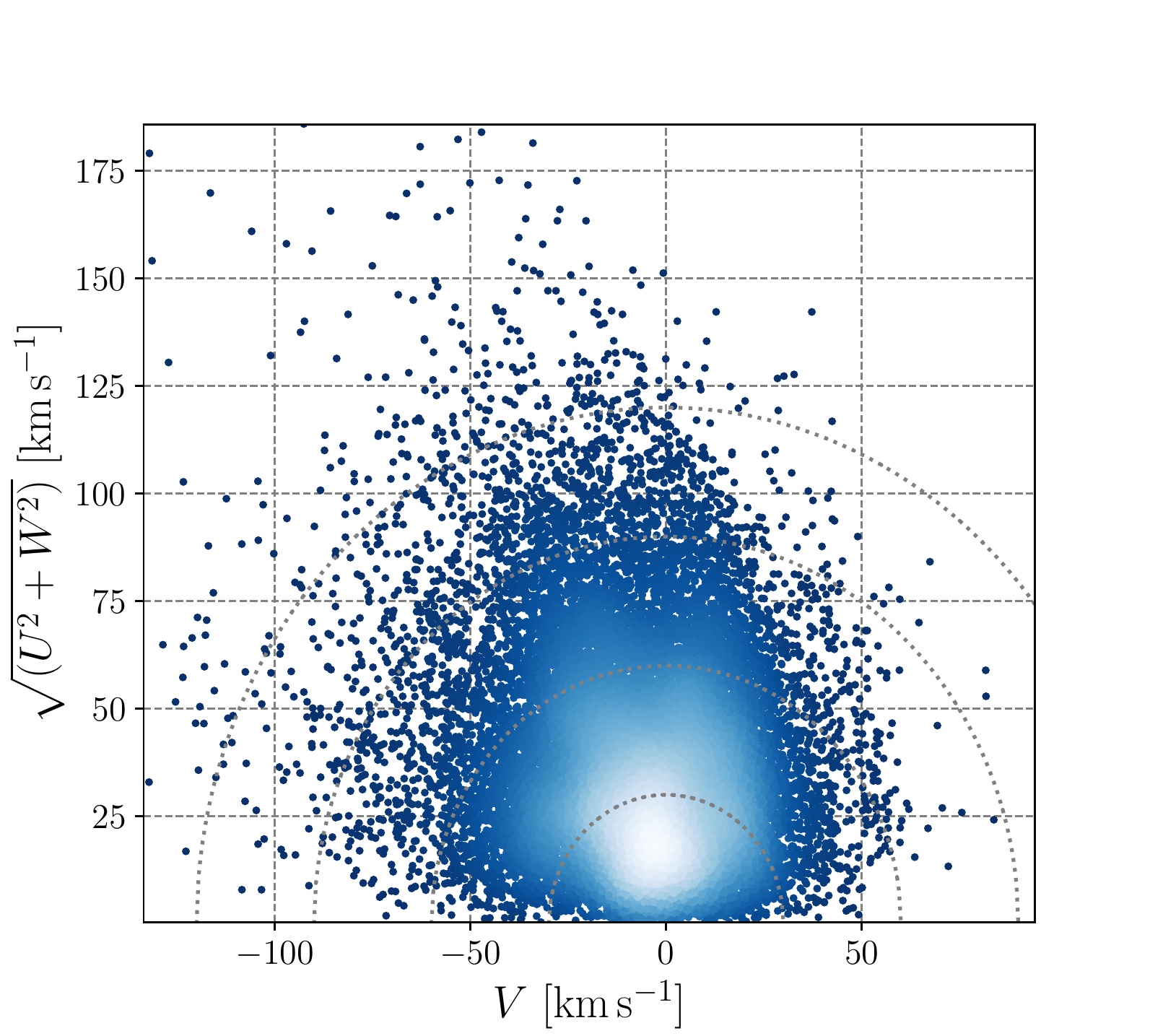}
\caption{Toomre diagram for the parent star sample. Dotted lines indicate constant total velocities $\vtot$ in steps of $30$, $60$, $90$ and $120\ \kms$. The color represents the density of points.}
\label{fig:Toomre}
\end{figure}

In Figure \ref{fig:Toomre} we show a plot of $\sqrt{U^2+W^2}$ vs. $V$ for our parent star sample. Such a plot is also known as a \textit{Toomre diagram}, and is considered a useful visual tool in distinguishing among the kinematic components of the Galaxy \citep[e.g.,][]{Bensby03}.

Many studies use the total space velocity of a star ($\vtot = \sqrt{U^2 + V^2 + W^2}$) combined with its metallicity as an indicator to its affiliation with a dynamic component of the Galaxy. It is customary to affiliate stars with $\vtot < 40\ \kms$ and $\feh > -0.2$ with the thin disk, while stars with $\vtot = 70$\,--\,$180\ \kms$ and $\feh < -0.3$ are usually identified as thick-disk stars \citep{Bensby14, Kordopatis15, DuongGALAH18}.

\subsection{Planet Occurrence Rates} 
\label{subsec:Occurrence}
We used the definition of occurrence rate as the average number of planets per star in a specific cell of stellar properties, and estimated it using the Inverse Detection Efficiency Methodology IDEM \citep{Foreman-Mackey14}, used by, e.g., \citet{petigura2018california}. The method accounts for the detection sensitivity -- the product of the pipeline detection efficiency $p_\mathrm{det}$ and the transit probability $p_\mathrm{tr}$. It uses completeness correction weights ($w=1/p=1/(p_\mathrm{tr}p_\mathrm{det})$), as listed in Table~4 of \citet{FultonPeti18}.

We have divided the $(\feh,\vtot)$ plane into cells, and estimated the planet occurrence in each cell $f^{i,j}$ by summing the completeness correction weights of all planets in the cell and normalizing by the number of stars in the cell $n_{*}^{i,j}$:
\begin{equation}
  f^{i,j}=\frac{1}{n_{*}^{i,j}} \sum_{k=1}^{n_\mathrm{pl}^{i,j}} w_k
    \label{eq:Occ}
\end{equation}
where $k$ labels each planet in the cell $(i,j)$, and $n_\mathrm{pl}^{i,j}$ is the number of planets in the cell. Following \citet{Hill18} we estimated the uncertainties of $f^{i,j}$ by:
\begin{equation}
  \Delta f^{i,j}=\frac{f^{i,j}}{\sqrt{n_\mathrm{pl}^{i,j}}} .
  \label{eq:error}
\end{equation}
We repeated this analysis for several different cell partitions, in order to make sure that the trends we report below are not due to the specific choice of partition.

\subsection{Stellar Kinematics and $\alpha$~Enhancement}
\label{subsec:kinematicsandalpha}
The main benefit of using the LAMOST catalog is the availability of LAMOST spectra for a large fraction of \textit{Kepler} stars. However, currently LAMOST data releases do not contain any estimate of $\alpha$~enhancement: $[\alpha/\mathrm{Fe}]$. Therefore, in order to study the dependence of planet occurrence on iron and $\alpha$-element abundances, we chose to use the stellar kinematics instead of $\alpha$~enhancement. As noted earlier, although stellar kinematics by itself is not an ideal indicator to classify disk sub-populations, it can still be used in combination with the iron abundance as a proxy for the mean $\alpha$~enhancement. 

To demonstrate this, we use the Galactic Archaeology with HERMES (GALAH) DR2 catalog \citep{BuderGALAH18}, which contains stellar parameters and abundances for up to $23$ elements, to present the expected relation between $\alpha$~content, iron abundance and kinematics.

\begin{figure}[ht!]
\epsscale{1.3}
\plotone{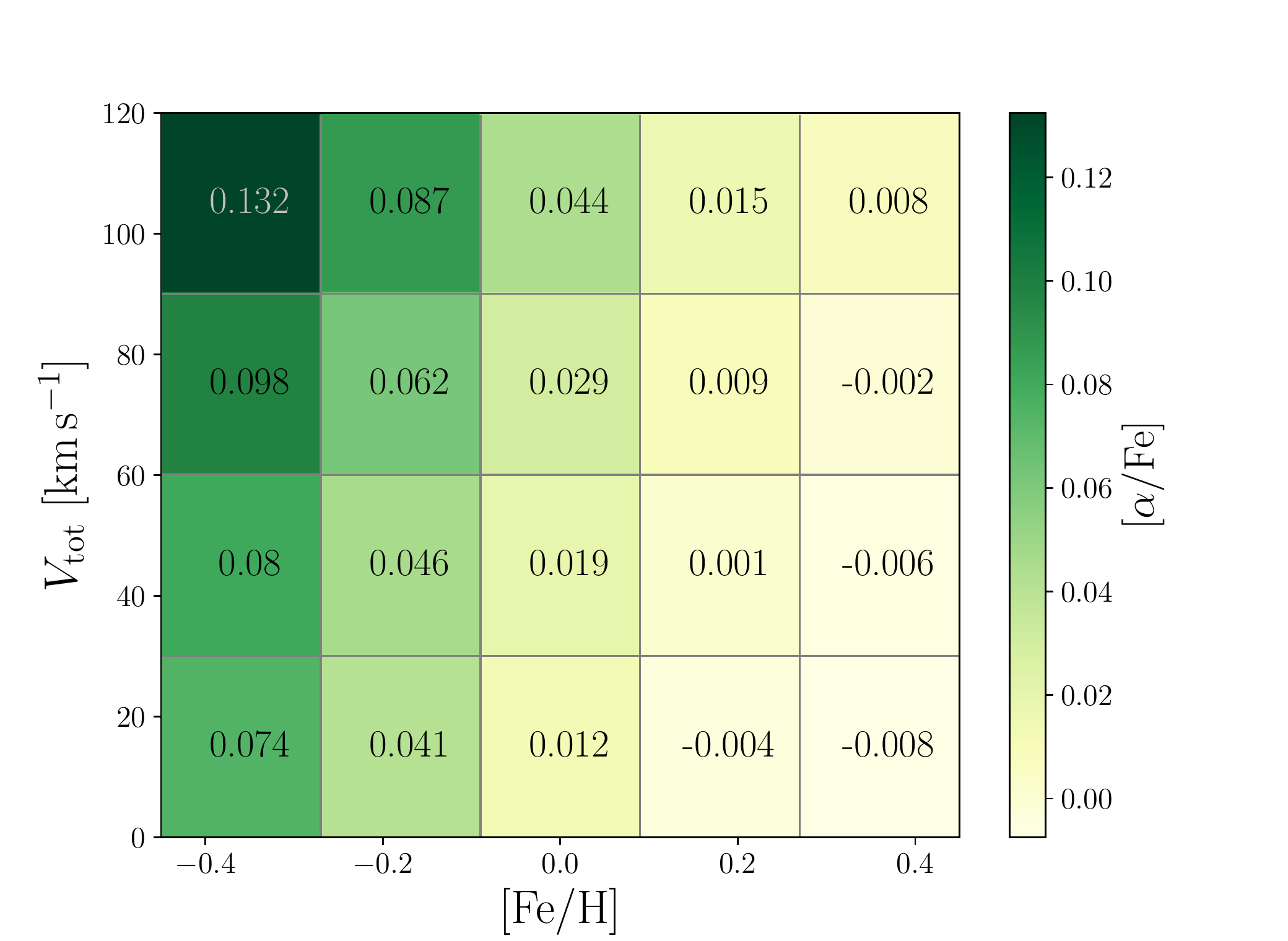}
\caption{Mean $\alpha$~enhancement in a binned diagram of stellar iron abundance and kinematics. The error in each bin is similar ($\sim 0.025$). As can be seen, there is a trend towards higher $\alpha$~content with higher kinematics and lower iron content. The data was acquired from the GALAH DR2.}
\label{fig:GALAH}
\end{figure}
By using similar filters and cuts as we have used for the LAMOST stellar sample, we compile a sample of  $45451$ stars with measured iron abundance, $\alpha$~enhancement and kinematics. We show in Figure \ref{fig:GALAH} the calculated average $\alpha$~content on a $5 \times 4$ grid of iron abundance ($-0.45 < \feh < +0.45$) and total velocity ($0 < \vtot < 120\ \kms$).  Although this does not allow us to classify the stars into their disk sub-populations, we do see a clear rising trend of the average $\alpha$~content as we move to lower iron abundance and higher total velocity.

\section{Results}
\label{sec:Results}
\begin{figure*}[ht!]
\epsscale{1.2}
\plotone{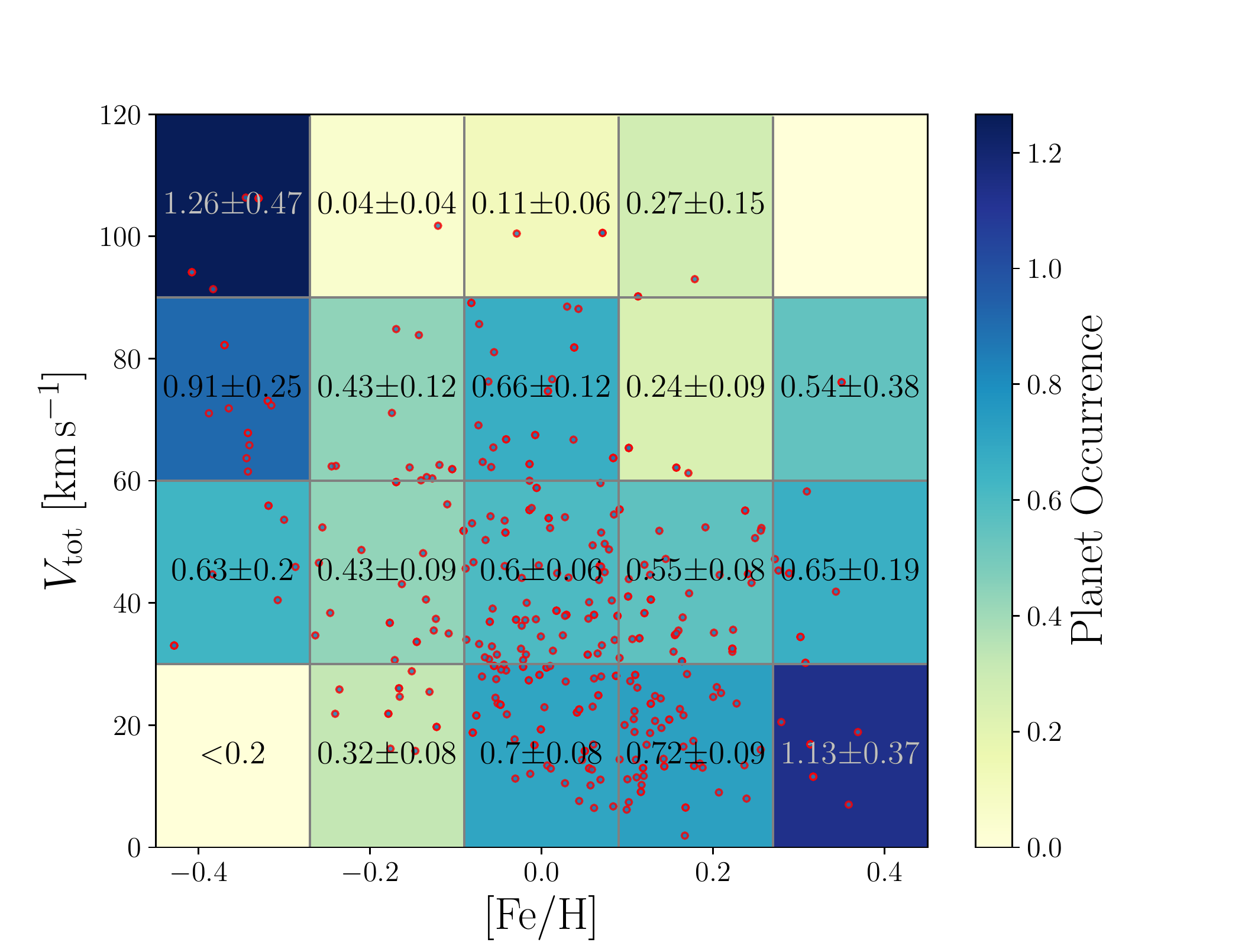}
\caption{Evenly-spaced grid of small close-in planet occurrence rates as a function of iron abundance ($-0.45 < \feh < 0.45$) and total velocity ($0 < \vtot < 120\ \kms$). Planet occurrence and uncertainty are given as the mean number of planets per star, and colored accordingly. The red dots represent the planet host stars in our sample.}
\label{fig:Occurrence}
\end{figure*}

Using the prescription detailed in Section \ref{subsec:Occurrence}, we have estimated the small close-in planet occurrence rate $f^{i,j}$ for a $5 \times 4$ evenly-spaced grid of iron abundance ($-0.45 < \feh < +0.45$) and total velocity ($0 < \vtot < 120\ \kms$)\footnote{We chose to exclude from our occurrence analysis a single planet host star (Kepler-1619, also labeled as KOI-4693), which seems to be an outlier with $\vtot= 148\ \kms$ and $\feh = -0.23$.}. We present our results in Figure \ref{fig:Occurrence}, where the cells are shaded and annotated according to the corresponding occurrence rate. As for the most bottom-left cell, since it does not include any planet host star, we have only estimated a rough upper limit. To do so, we assumed a detection of a single planetary system with the same number of planets as the average number of planets per planetary system in our complete sample $n_\mathrm{pl}=N_\mathrm{p}/N_\mathrm{ps}$, where $N_\mathrm{p}$ and $N_\mathrm{ps}$ are our total number of KOIs and planet host stars respectively. We then calculated the average planetary weight $\overline{w}$ and used Equation \ref{eq:Occ} to obtain an upper-limit occurrence rate.

An examination of Figure \ref{fig:Occurrence} reveals two different trends of rising small close-in planet occurrence rate. First, a rise with total velocity at the lowest iron abundance bin: from an almost negligible planet occurrence rate around slow stars ($\vtot < 30\ \kms$), to a rate of $\sim 1.2$ planets per star at the highest velocities. The second rising trend is with iron abundance, which is evident only at the lowest-velocity regime ($\vtot < 30\ \kms$), up to a rate of $\sim 1.1$ planet per star in the iron-rich regime. No clear trend can be noticed in other parts of the Figure.

\section{Discussion}
\label{sec:Discussion}
 Our results suggest there are two favored regions in the $(\feh,[\alpha/\mathrm{Fe}])$ plane where stars tend to better form and maintain small close-in planets. Currently, since $\alpha$~content information is not listed for a large portion of KIC stars, we chose to use kinematics as a proxy for $\alpha$~content. It is important to note that the actual relation we observed is between iron content and stellar kinematics (characterized by the total velocity). At present, it is unclear whether there are other quantities (except of $\alpha$-enhancement) that might also affect planet occurrence rate in regions of iron-poor fast stars. For example, stellar age or eccentric galactic orbit are also linked to $\alpha$-rich fast stars. 
 
 Although the use of kinematics as a proxy to $\alpha$-enhancement might lead to some contamination of our database with slow $\alpha$-rich stars (or fast $\alpha$-poor), the general trends of rising small-planet occurrence rate with: 1) iron-poor higher velocities ($\alpha$~enhanced) stars, 2) iron-rich, low velocities stars, is suggestive. Furthermore, although the metallicity of fast iron-poor stars is much lower than that of slow iron-rich stars, the small-planet formation efficiency seems to be similar. In fact, a future analysis of KOIs occurrence rate on a $(\feh,[\alpha/\mathrm{Fe}])$ diagram might even suggest a higher relative occurrence of planets around the iron-poor $\alpha$-rich stars. This, may shed light on the contribution of $\alpha$-enhancement to the higher profusion of small close-in planets among thick disk stars. 

In the literature there is some ambiguity regarding the dependence of small close-in planet occurrence rate on the iron abundance. While \citet{petigura2018california} suggest that the rate does not depend strongly on the iron abundance, \citet{BrewerMulti18} identify two iron-abundance regions (high and low) of high occurrence rate. Figure \ref{fig:Occurrence} may point to the missing factor: the additional dimension of $\vtot$. Based on the relation between $\vtot$ and $\alpha$ enhancement for iron-poor stars (Section \ref{subsec:kinematicsandalpha}), this may hint at a rising occurrence rate at iron-poor, $\alpha$-rich stars. We argue that an analysis of the dependence of planet occurrence rate on overall metal content is misleading without including additional information on both iron and $\alpha$ element abundances. This additional information may put the analysis in the galactic context of the various galactic components \citep[e.g.,][]{Bensby14, Kordopatis15}.

Although both trends suggest a higher frequency of small close-in planets in more metal-rich regions (in iron or $\alpha$~elements) as implied by the core-accretion model, the fact that these suggestive trends are absent in the other regions is intriguing. On the one hand, one can argue that this result supports \cite{Buchhave12} conclusion that small planets are found in a wide range of metallicities. On the other hand, as was already suggested by \cite{Gonzalez2009} and \cite{Adibekyan12}, although small planets can be found in iron-poor regimes, they are mostly enhanced by other metals. Our current work does not allow us to conclusively settle this conundrum. Moreover, theories of planetesimals forming by dust condensation and coagulation to pebbles are still struggling to explain small planet formation \citep{Safronov69, Nimmo2018}. Likewise, our current results do not allow us to further state with confidence whether small-planet formation efficiency and survivability is higher in $\alpha$-rich stars than in iron-rich stars. The relatively low number of planet candidates around iron-poor stars does not allow us to extend our analysis to study other even less populated subgroups such as: hot small planets ( $P<10\ \mathrm{days}$), super-Earth ($\rp< 1.7\ R_\earth$) or sub-Neptunes ($\rp = 1.7$\,--\,$4\ R_\earth$). With the next release of LAMOST DR5, the LAMOST-KIC sample will be almost doubled \citep{ZongLK2018}, potentially allowing us to analyze planet occurrence for a larger set of KOIs, and maybe improve our estimates of planet occurrence rate. Furthermore, a careful analysis of the LAMOST-\textit{Kepler} spectra to estimate $\alpha$~content would be of high importance. Despite the low resolution of LAMOST spectra, new generation of data-driven methods of machine learning are currently offering the opportunity to better estimate elemental abundances \citep{Xiang17, LeungBovy18}. However, current results are still limited to only a small fraction of the LAMOST stars, especially the evolved ones \citep{Ho17, Boeche18}. Furthermore, disagreement between the chemical abundances derived by different spectral synthesis techniques also cause some ambiguity \citep{Hinkel16}. 

Although not part of our main analysis, we can nevertheless make some remarks concerning the occurrence rate of close-in giant planets. From a total of $22$ giant planets ($\rp = 4$\,--\,$14\ R_\earth$) in close orbit ($P<100\ \mathrm{days}$), we did see, as expected, a clear rising trend with iron abundance for the slowest stars ($\vtot < 30\,\kms$): from $\la 0.86\ \%$ at $\feh < -0.1$ to $\sim 24\ \%$ for $\feh > +0.25$. However, we saw no evidence for the positive trend we had detected with $\vtot$ for small planets. These results are also consistent with previous works \citep[e.g.,][]{Maldonado18, BrewerMulti18}, suggesting that stars with low iron content (no matter their $\alpha$~content) do not seem to form or maintain close-in giant planets. More distant giant planets (not included in our survey) do seem to be more common around stars of low-iron and high~$\alpha$ content, as the latest RV surveys for metal-poor stars suggest \citep{Maldonado18, Barbato19}. 

The occurrence rate we estimated pertained to the average number of planets per star. This is the final outcome of both planet formation and dynamical evolution. In our analysis as well as in most other planetary occurrence studies, there should be another correction for the weights of multi-planetary systems (e.g., the effect on planet occurrence should distinguish between a star with four orbiting planets and four different stars each having one planet). In future work, it should be interesting to find the occurrence rate of planet-host stars, disregarding the multiplicity. As recently noted by \cite{Zhu19}, this is not a quantity which can be easily derived, in particular for transit surveys such as \kep. This will further allow us to calculate the average multiplicity: the average number of planets per planetary system (not per star). Studying both occurrence rate and multiplicity may help to disentangle the effects of planet formation and dynamical evolution.  

The recently launched TESS mission \citep{Ricker15} and upcoming missions such as PLATO \citep{Rauer14} will certainly increase the available sample of small planets and thus potentially will probe a larger sample of the less-common iron-poor stars. Hopefully, they will allow a better understanding of the intriguing relation we suggest in this paper.

\acknowledgments

This research was supported by the ISRAEL SCIENCE FOUNDATION (grant No. 848/16) and by the Ministry of Science, Technology and Space, Israel. This work includes data collected by the \textit{Kepler} mission. Funding for the \textit{Kepler} mission is provided by the NASA Science Mission directorate. This work has also made use of data from the European Space Agency (ESA) mission \gaia\ (\url{https://www.cosmos.esa.int/gaia}), processed by the \gaia\ Data Processing and Analysis Consortium (DPAC, \url{https://www.cosmos.esa.int/web/gaia/dpac/consortium}). Funding for DPAC has been provided by national institutions, in particular the institutions participating in the \gaia\ Multilateral Agreement. This paper also uses data from the LAMOST survey: Guoshoujing Telescope (the Large Sky Area Multi-Object Fiber Spectroscopic Telescope LAMOST) is a National Major Scientific Project built by the Chinese Academy of Sciences. Funding for the project has been provided by the National Development and Reform Commission. LAMOST is operated and managed by the National Astronomical Observatories, Chinese Academy of Sciences. 
The GALAH survey is based on observations made at the Australian Astronomical Observatory, under programmes A/2013B/13, A/2014A/25, A/2015A/19, A/2017A/18. We acknowledge the traditional owners of the land on which the AAT stands, the Gamilaraay people, and pay our respects to elders past and present.
This paper includes data that has been provided by AAO Data Central (\url{datacentral.aao.gov.au}).

\facilities{LAMOST, Kepler, Gaia, AAT (HERMES)}

\end{document}